\documentstyle[11pt,newpasp,twoside]{article}
\index{supernova remnant!Puppis A}
\index{supernova remnant!RCW 103}
\index{supernova remnant!Vela Jr}
\index{supernova remnant!G347.3--0.5}
\index{central compact objects}

\def\HI{\rm H\,{\sc i}}
\def\mjb{mJy beam$^{-1}$}

\def\k{km s$^{-1}$}
\def\pp{^{\prime\prime}}
\def\cm2{cm$^{-2}$}

\def\edcomment#1{\iffalse\marginpar{\raggedright\sl#1\/}\else\relax\fi}
\reversemarginpar

\begin{document}
\title{Imprints of neutron stars in the interstellar medium}
 \author{Estela M. Reynoso}
\affil{Instituto de Astronom\'\i a y F\'\i sica del Espacio (IAFE),
c.c. 67, Suc. 28, 1428 Buenos Aires, Argentina}
\affil{and School of Physics, 
A29, University of Sydney, Camperdown Campus, 2006 NSW, Australia} 
\author{Simon Johnston, Anne J. Green}
\affil{School of Physics, 
A29, University of Sydney, Camperdown Campus, 2006 NSW, Australia}
\author{W. M. Goss}
\affil{National Radio Astronomy Observatory, PO Box 0, Socorro, NM 87801, USA}
\author{Gloria M. Dubner, Elsa B. Giacani}
\affil{IAFE, c.c. 67, Suc. 28, 1428 Buenos Aires, Argentina}

\begin{abstract}
We have carried out an \HI \ survey towards 
X-ray central compact objects\- (CCOs)\- inside
supernova remnants (SNRs) which shows that many of them are placed within
local \HI \ minima. The nature of these minima is not clear, but the most
likely explanation is that the CCOs have evacuated the neighboring gas. 
This survey also allowed us to detect a weak, diffuse radio nebula 
inside the SNR Vela Jr, probably created by the winds of its associated CCO.
\end{abstract}

\section{Introduction and Observations}
Several X-ray point sources with no radio counterpart, generically called CCOs,
have recently been detected near the center of SNRs. In most cases, these 
sources are claimed to be the neutron stars (NSs) left behind after the 
supernova explosions. One of them, 1E 1207.4-5209, was found to lie at the 
center of an \HI \ depression (Giacani et al. 2000), raising the question 
of whether the hot atmosphere of the NS is capable of heating up the 
neighboring gas and producing the observed depression.
We present the results of a search for similar traces in the 
interstellar gas towards a sample of CCOs in southern SNRs.

The environs of the CCOs RX J0822--4300 in Puppis A, 1E 161348--5055 in RCW 
103, CXOU J08521.4--461753 in Vela Jr and 1 WGA J1713.4--3949 in G347.3--0.5
were observed in the $\lambda21$ cm \HI \ line and in continuum using 
the Australia Telescope Compact Array. The data were combined with single
dish data from the Parkes telescope (McClure-Griffiths et al. 2001). 

\section{Results and Discussion}
RX J0822--4300, the CCO in Puppis A, was found to lie between two opposite 
lobe-like \HI \ minima (Reynoso et al. 2003a). The lobes are aligned with the 
proper motion of the CCO, assuming that the explosion site of the supernova is
given by the optical expansion center measured by Winkler et al. (1988). The 
lobes are centered at the same systemic velocity previously measured for Puppis 
A, +16 \k. Reynoso et al. (2003a) propose that this \HI \ structure is created 
by the ejection of two opposite jets from the CCO. More X-ray observations 
towards RX J0822--4300 are needed to search for jets (like in Vela or the Crab 
pulsar) and measure the CCO's proper motion.

At  +3 \k, the \HI \ shows another depression coincident with RX J0822--4300,
but in this case the morphology and size are similar to the minimum
found by Giacani et al. (2000) around the CCO associated with G296.5+10.0. 
RCW 103 represents the third case in which the associated CCO has created 
an \HI \ depression (Reynoso et al. 2003b). In all three cases, these
\HI \ features have sizes of 1 to 3 pc, the CCOs are off-centered
by $\sim 0.3$ pc, and the missing masses are estimated to be 0.1 to 0.3 
$M_\odot$. It is very unlikely that these depressions are due to
self-absorption, since the involved temperatures should be $\sim 100$ K, 
too low for a SNR interior. Most probably, the CCOs swept up the surrounding
gas (Reynoso et al. 2003b). In all cases, the measured \HI \ column densities 
favor blackbody rather than power law fits to the X-ray spectra.

For G347.3--0.5, a preliminary analysis of our data did not allow
us to find any feature suggestive of being associated with the CCO down to a
limit of $\Delta T=5$ K (3$\sigma$). At a velocity compatible with the distance 
proposed to this SNR, there appears a tiny depression, marginally enclosing the 
CCO, but it does not look different than many other spots in the image.

Finally, the radio continuum data towards the SNR Vela Jr reveal an
elongated nebula, approximately $30^\prime$ in length and $14^\prime$ in
width, centered at the position of the CCO.
In addition, a compact source is found at the 
location of the CCO. The flux of this compact source is 7.2 \mjb, and its
size, $\sim 85\pp \times 27\pp$. Polarization and spectral index studies
will provide information to confirm if this emission arises from the pulsar
wind nebula created by CXOU J085201.4--461753. Such observations are
proposed early in 2004.

\end{document}